\documentclass{tMOP2e}

\usepackage{caption}
\usepackage{subcaption}
\theoremstyle{plain}

\theoremstyle{definition}

\theoremstyle{remark}

\begin{document}



\title{Suppression of the four-wave mixing amplification via  Raman absorption}

\author{
\name{Gleb Romanov\textsuperscript{a}, Chris O'Brien\textsuperscript{b} and I. Novikova\textsuperscript{a}$^{\ast}$\thanks{$^\ast$Corresponding author. Email: inovikova@physics.wm.edu}
}
\affil{\textsuperscript{a}The College of William \& Mary, Williamsburg VA 23185 USA \\
\textsuperscript{b} IQST and Department of Physics and Astronomy, University of Calgary, Calgary AB T2N 1N4, Canada
}
\received{\today}
}

\maketitle

\begin{abstract}

We propose a method to controllably suppress the effect of the four-wave mixing caused by the coupling of the strong control optical field to both optical transitions in the $\Lambda$ system under the conditions of electromagnetically induced transparency. At sufficiently high atomic density, this process leads to amplification of a weak optical signal field, that is detrimental for the fidelity of any EIT-based quantum information applications. 
Here we show that an additional absorption resonance centered around the idler field frequency, generated in such a four-wave mixing process, may efficiently suppress the unwanted signal amplification without affecting properties of the  EIT interaction.
We discuss the possibility of creating such tunable absorption using two-photon Raman absorption resonances in the other Rb isotope, and present some preliminary experimental results.    
\end{abstract}


\section{Introduction}

Realizations of strong coupling between an optical field and an ensemble of atoms using  two-photon processes -- such as electromagnetically induced transparency (EIT)~\cite{harris'97pt,marangos'98,lukinRMP03,fleishhauerLukinPRL00,fleishhauerLukinPRA02,novikovaLPR12} and off-resonant Raman interaction~\cite{polzikRMP10,euromemory,ReimPRL11} -- offer a tantalizingly simple method for all-optical quantum control, required for many quantum information applications~\cite{hilleryPRA00,stroudPRA06,DLCZ,polzik_book,kimbleNature08,lvovskyNPh09,polzikRMP10,euromemory,Walmsley01052015} and quantum sensor technologies~\cite{bachor_guide_2004,mitchel2010prl_sub_projection_noise,mikhailov2012sq_magnetometer}.
Many of the existing optical quantum memory protocols rely on such strong coupling to realize the reversible mapping between quantum states of light and long-lived atomic spin states.
In the last two decades there have been a number of proposals and proof-of-principle demonstrations for EIT- and Raman-based quantum memories for optical pulse propagating through atomic (and atom-like) media, as well as demonstrations of generation of entanglement between optical fields and atomic ensembles, en route to realization of quantum repeaters~\cite{DLCZ,GizinRepeatersRMP11}, on-demand single-photon sources, optical buffers, etc.

\begin{figure}
    \centering
    \includegraphics[height=2.5in]{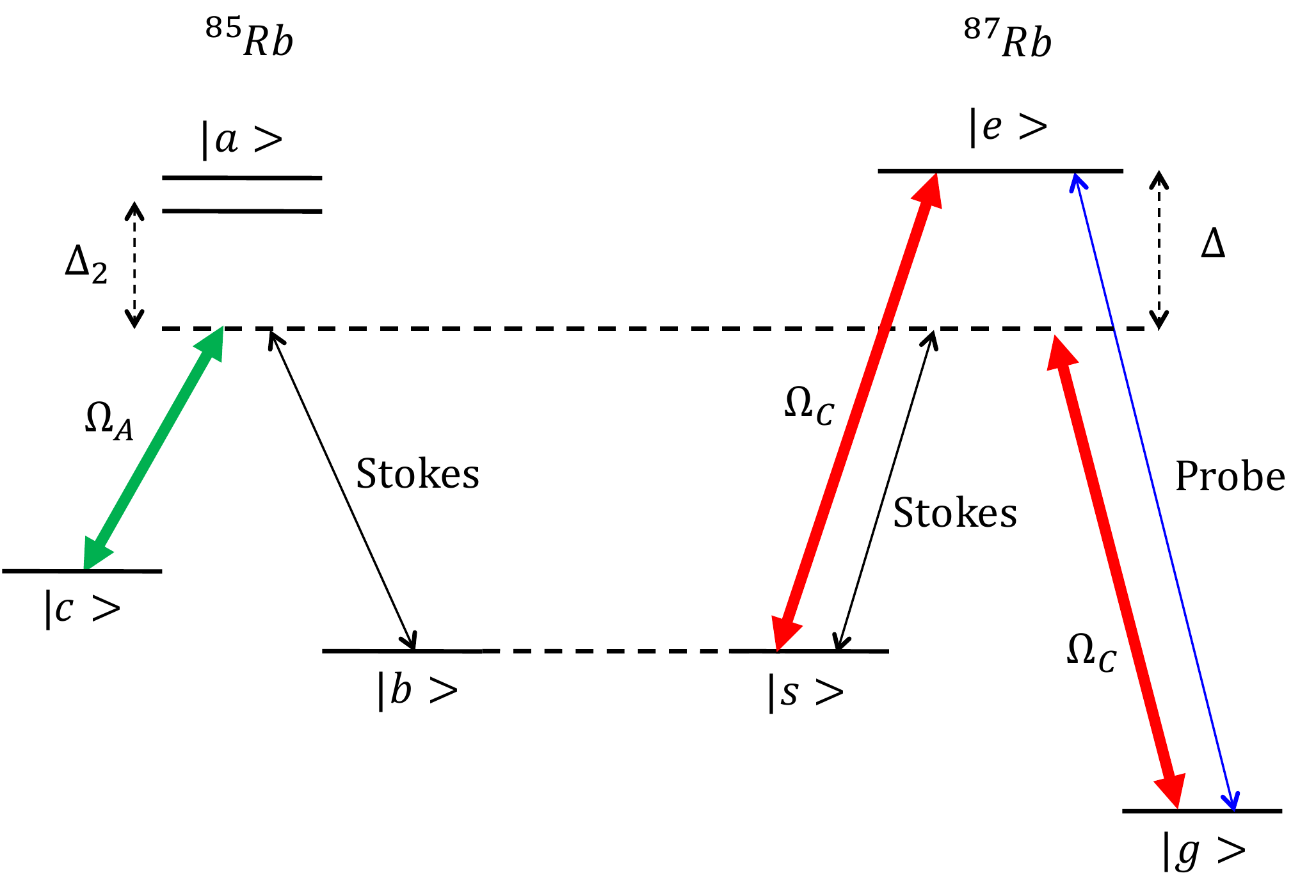}
    \caption{Diagram showing relevant laser fields and energy levels for two atomic systems. A weak probe (blue) and a strong control (red) interact with the first type of atoms under the resonant EIT conditions. The additional off-resonant coupling of the control field to the $|g\rangle - |e\rangle$ transition creates four-wave mixing conditions in a resulting double-$\Lambda$ system and leads to generation of the idler (Stokes) field (black). A second strong Raman control field (green) is tuned to induce a Raman absorption resonance for the Stokes field using the second type of atoms.
   }
    \label{fig:rb_levels}
\end{figure}

Both EIT and Raman effect rely on the strong coupling of an optical signal field and collective long-lived ensemble of atomic spins  by means of a strong classical optical control field in a $\Lambda$ configuration~\cite{marangos'98,lukinRMP03,fleishhauerLukinPRL00,fleishhauerLukinPRA02}.
Since the strength of such interaction is proportional to the number of atoms, the optimal performance often requires operation in high optical depth regime.
Unfortunately, the increasing optical depth of atomic ensemble also leads to effective enhancement of other nonlinear light-atom interactions, that may interfere with the expected performance.
For example, one is no longer allowed to disregard the off-resonant coupling of the control field with the other optical transition, giving rise to a double-$\Lambda$ system, shown in Fig. ~\ref{fig:rb_levels} and resulting in amplification of the optical signal field and generation of an additional Stokes field~\cite{lukin97prl,narducciPRA04,HaradaPRA08,hong09,Howell_2009,phillipsPRA11,gengNJP2014}.
While such process are very promising for generation of two-mode squeezing and entanglement~\cite{shahriarOC98,lettPRA08,lettSci08}, it poses a serious problem for quantum memory operation, as it creates uncorrelated photons in the signal field channel~\cite{laukPRA13,MichelbergerNJP15}.
Several approaches has been proposed to suppress four-wave mixing by optimizing  frequencies ~\cite{HaradaPRA08,bashkanskyPRA13} or polarizations~\cite{zhangPRA14}; however, the potential disadvantage of these methods that the required operational parameters may not correspond to the optimal memory performance.  

Here we discuss the possibility to control four-wave mixing in a three-level system without deteriorating the coherent properties of EIT by introducing  an additional absober resonant exclusively with the idler (Stokes) field.
While this cannot stop idler photons from being created, FWM is a stimulated process and by removing idler photons its efficiency is greatly decreased.
We demonstrate that in theory it is possible to suppress the four-wave mixing gain in the signal channel with sufficient Stokes absorption.
That makes it a plausible avenue for improving the quantum memory fidelity by suppressing FWM-induced excess noise.  

The main challenge for the experimental realization of this proposal is to create an absorption resonance with required parameters for the Stokes field without affecting the rest of the system. 
We analyze one such possibility that takes advantage of existence of two stable Rb isotopes with very similar resonance frequencies: one of them can be used for realization of the quantum memory, while the other - for Stokes absorption.
The two-photon resonance from a far-detuned Raman system can be used to create an effective two level absorption which can be tailored to absorb the FWM-generated Stokes field \cite{mikhailov04josab,OBrien_RIC1}. 

The paper is organized as follows.
In Section~\ref{simpletheory} we provide a general analysis of the effect of added Stokes absorption in a three-level system, and discuss its possible realization in Rb vapor.
The experimental setup is described in Section~\ref{expsetup}, and preliminary experimental results are presented and analyzed in Section~\ref{results}.
Finally, a discussion of the optimal setup is given in Section~\ref{discussion}.

\section{Theory}\label{simpletheory}

The theoretical description of the resonant four-wave mixing in a double-$\Lambda$ configuration has been already developed in previous works~\cite{hong09,Howell_2009,phillipsPRA11,laukPRA13}. 
Following the treatment in \cite{hong09,laukPRA13} for the double-$\Lambda$ scheme in the approximation of negligible spin coherence relaxation rate ($\gamma_{gs} = 0$, ), the output fields can approximately be expressed for resonant fields
\begin{align}
& \hat a ^{out} _S  =\cosh (\frac{D \gamma_{ge}}{\Delta})\hat a _S ^{in} +   i\sinh (\frac{D \gamma_{ge}}{\Delta})(\hat a _I ^{in})^\dag \label{eq:1} \\
& (\hat a ^{out} _I)^\dag = - i \sinh (\frac{D \gamma_{ge}}{\Delta}) \hat a _S ^{in} +  \cosh (\frac{D \gamma_{ge}}{\Delta}) (\hat a _I ^{in})^\dag, \label{eq:2}
\end{align} 

where $\hat a _S$ and $\hat a _I$ are the destruction operators for the signal field and the Stokes field, correspondingly.
Here $D$ is the optical depth of the atomic medium, $\Delta$ is the detuning of the control field from the $|g\rangle - |e\rangle$ atomic transition (that forms the second $\Lambda$ link together with the generated Stokes field), and $\gamma_{ge}$ is the optical decoherence rate.
The expression for the output signal clearly indicates two effects due to FWM.
First, the presence of the $ (\hat a _I ^{in})^\dag$ term in Eq.(\ref{eq:1}) describes amplification of the output signal field if there is an input idler field.
Second, even if there is no input idler field, a number of photons proportional to $|\sinh (D \gamma_{ge}/\Delta)|^2$ are created through FWM.
Such amplification is detrimental to the fidelity of an optical quantum memory based on EIT~\cite{laukPRA13} and Raman resonances~\cite{MichelbergerNJP15}, because each of these additional photons does not depend on the input signal and effectively adds uncorrelated noise to the optical signal. 

If loss is introduced for the idler field, then the efficiency of FWM can be reduced.
We first consider a general case of a hypothetical absorber with an optical depth  $D_{\text{abs}}$ resonant only with the idler field, with no effect on the other optical fields. 
We start with the Maxwell-Bloch equations:
\begin{align}
i\partial _t \hat \sigma _{ge} &= -i\gamma_{ge} \hat \sigma_{ge} -g\hat a _S - \Omega_c \hat \sigma _{gs}, \label{eq_1} \\
i\partial _t \hat \sigma _{gs} &= -g\frac{\Omega_c}{\Delta}\hat a _I ^\dag - \Omega_c ^* \hat \sigma _{ge}, \label{eq_2} \\
(\partial _t + c \partial _z) \hat a _S &= igN \hat \sigma _{ge}, \label{eq_3} \\
(\partial _t + c \partial _z) \hat a _I ^\dag &= -igN \frac{\Omega_c^*}{\Delta} \hat \sigma _{gs} - \frac{c}{L} D_{\text{abs}} \hat a _I. \label{eq_4}
\end{align}
where $\Omega_c$ is the control field Rabi frequency, $N$ is the number of atoms, and $g$ is the single photon Rabi frequency for the EIT transition, $c$ is the speed of light, and $L$ is the length of both EIT and absorbing media.  
Under the approximation of the slowly-varying coherences, we can solve Eq.(\ref{eq_1}) and Eq.(\ref{eq_2}) adiabatically and, plugging the results into Eqs.(\ref{eq_3},\ref{eq_4}), we get two equations for the signal and idler field operators:
\begin{align}
\frac{\partial}{\partial z} \hat a _S (z) &= -i \frac{g^2 N}{c\gamma_{ge}} \frac{\gamma _{ge}}{\Delta} \hat a _I ^\dag (z), \label{eq_5} \\
\frac{\partial}{\partial z} \hat a_I ^\dag (z) &= \frac{g^2 N}{c\gamma_{ge}} \frac{\gamma _{ge} ^2}{\Delta^2} \hat a _I ^\dag (z) + i\frac{g^2N}{c \gamma_{ge}} \frac{\gamma_{ge}}{\Delta} \hat a _S (z) - \frac{D_{\text{abs}}}{L} \hat a _I ^\dag (z). \label{eq_6}
\end{align}

For two-level absorption in resonance with the idler field, and optical depth $D_{\text{abs}}$, we can solve Eqs.(\ref{eq_5},\ref{eq_6}) then if we assume $\gamma_{ge} \ll \Delta$ and $D_{abs} \Delta \gg D\gamma_{ge}$, the output becomes:
\begin{align}
\hat a ^{out} _S  = \hat a _S ^{in} e^{D\frac{\gamma_{ge}^2}{\Delta^2}\frac{D}{D_{\text{abs}}}} - \frac{i}{2}\frac{D}{D_{\text{abs}}}\frac{\gamma _{ge}}{\Delta} (\hat a _I ^{in})^\dag e^{D\frac{\gamma_{ge}^2}{\Delta^2}\frac{D}{D_{\text{abs}}}}.
\end{align}
where the optical depth of the EIT medium is defined as 
\begin{align}
D = \frac{g^2 N L}{c \gamma _{ge}}.
\end{align}
Thus, if the loss we introduced is larger than the FWM gain, it significantly reduced the effects of FWM. 
Now we can find the ratio of noise photons $N_{\text{abs}}$ created with the additional absorption to the number of noise photons $N_{\text{FWM}}$ created without absorption taken from
\cite{laukPRA13},
\begin{align}
\frac{N_{\text{abs}}}{N_{\text{FWM}}} = \frac{D^2}{D_{\text{abs}}^2} \frac{\gamma _{ge} ^2}{\Delta ^2} e^{-2D\frac{\gamma _{ge}}{\Delta} \left( 1 - \frac{\gamma _{ge}}{\Delta} \frac{D}{D_{\text{abs}}} \right)}. \label{eq:11}
\end{align}
It is easy to see that with sufficient Stokes absorption the number of noise photons in the signal field can be greatly decreased.

\begin{figure}
   \centering
   \begin{subfigure}[b]{0.45\textwidth}
       \includegraphics[width=\textwidth]{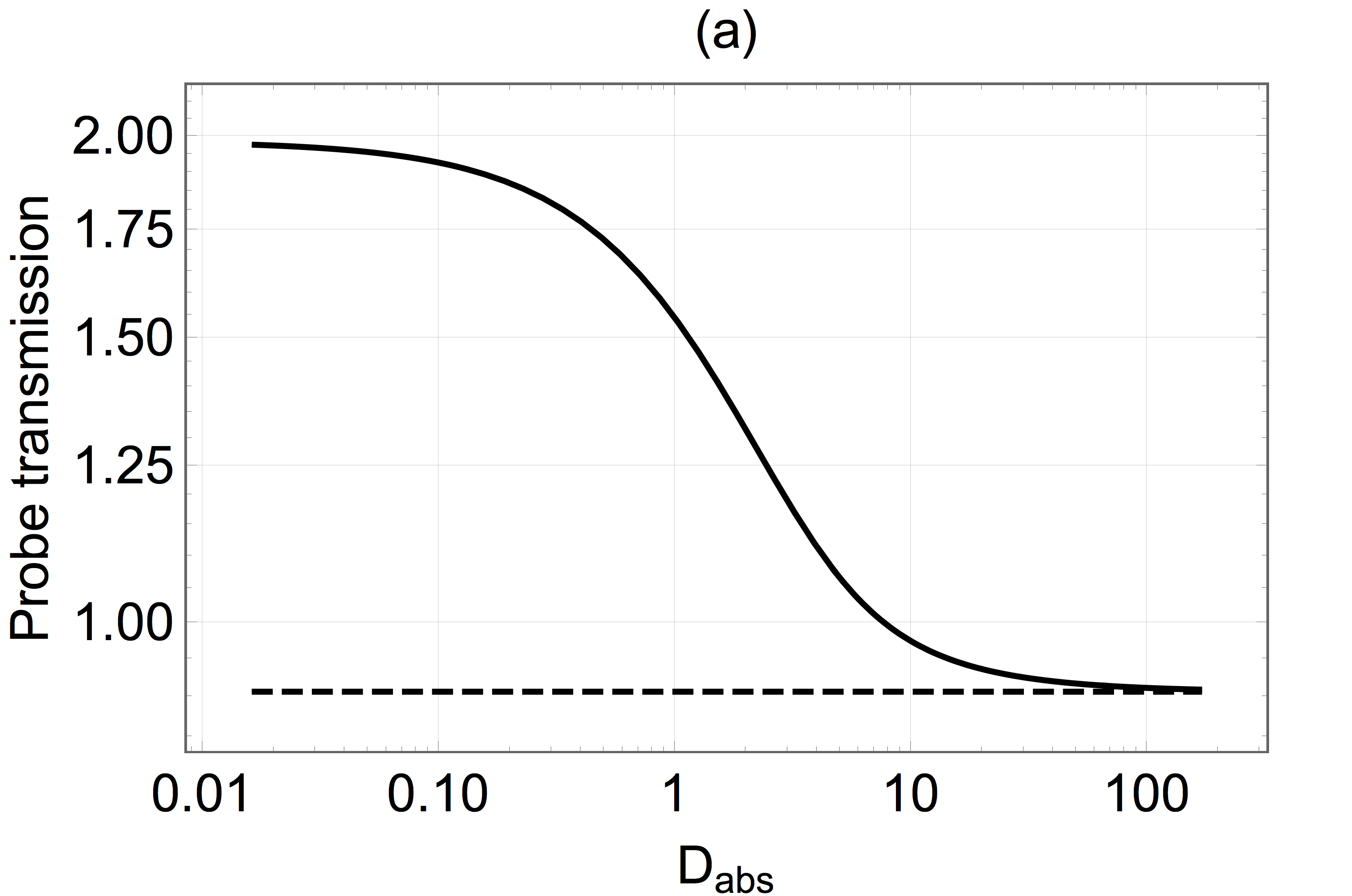}
       \label{fig:fwm_reduction_probe_vs_absorption}
   \end{subfigure}
   \begin{subfigure}[b]{0.45\textwidth}
       \includegraphics[width=\textwidth]{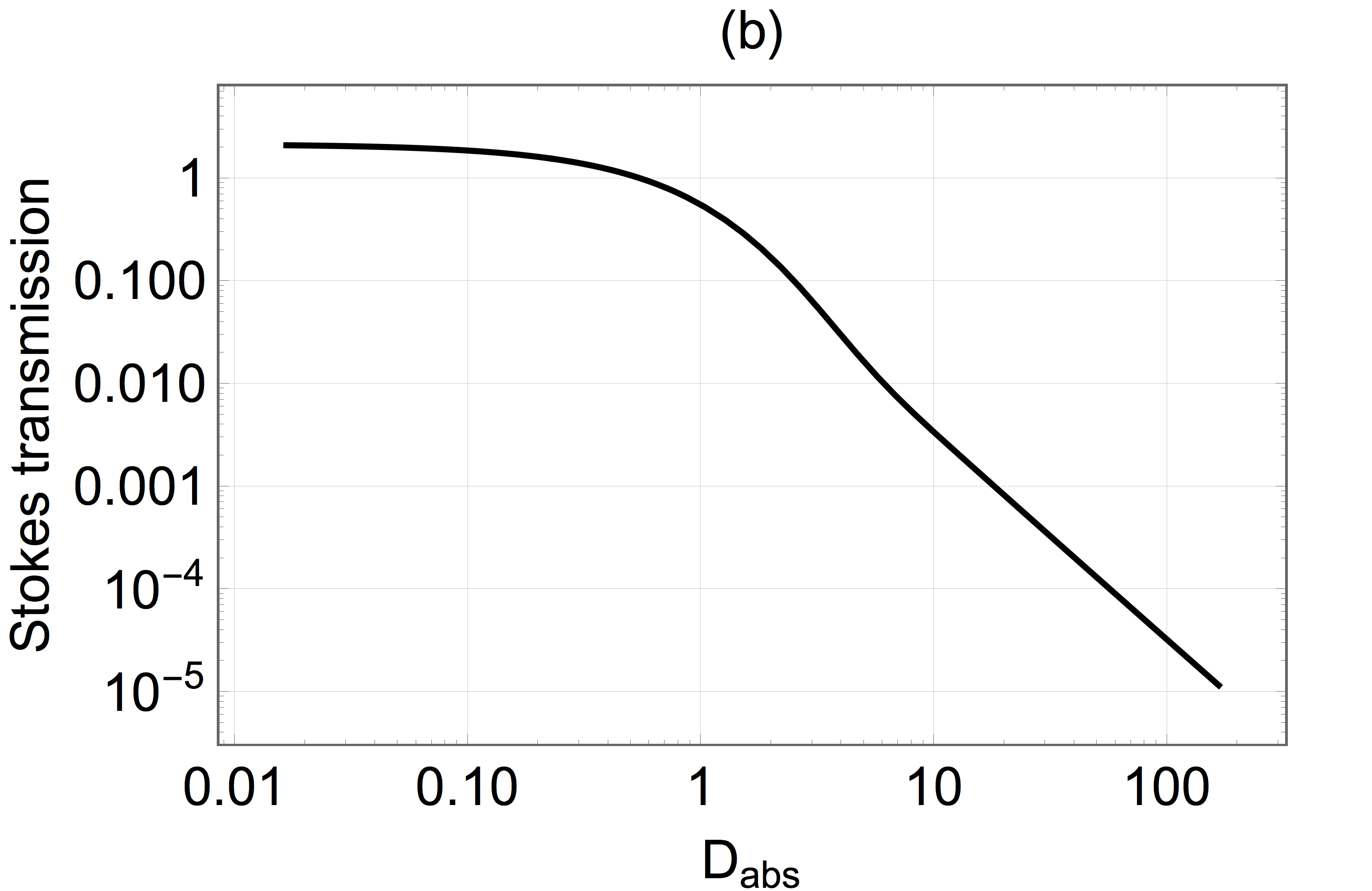}
       \label{fig:fwm_reduction_stokes_vs_absorption}
   \end{subfigure}
   \caption{Suppression of the four-wave mixing gain in probe (a) and Stokes (b) channels as a function of $D_{abs}$.
   The dashed line on the probe graph indicates maximum EIT transmission with no FWM.
    The parameters of the EIT atomic system are chosen to represent the optical transitions of the D${}_{1}$ line of ${}^{87}$Rb (see discussion in the text below).}
    \label{fig:fwm_reduction_vs_absorption}
\end{figure}

The effectiveness of the proposed method is illustrated in Fig.~\ref{fig:fwm_reduction_vs_absorption}, that depicts maximum amplitudes for the probe and Stokes fields for different values of $D_{abs}$. We chose the parameters of the atomic system such that in the absence of the Stokes absorption there is a significant amplification for the probe field, such that the output probe field intensity doubled compare to its input value. Noticeable Stokes field was generated at the output of the cell as well, as expected in the four-wave mixing process. As we increased the value of the optical depth for the resonant Stokes absorption, we observed  suppression of the Stokes field generation. Simultaneously, the output power of the signal field decreased to its value expected from the pure EIT propagation under these conditions ($\approx 95\%$), which shows that the coherent properties in the EIT system were not affected by our manipulations with the four-wave mixing channel.   

Unfortunately, there is no two-level atomic system that is in exact resonance with the generated idler field. 
Instead, we suggest introducing tunable absorption resonances in a far-detuned three-level $\Lambda$-system using a different atomic species, as was done for the refractive index control \cite{yavusPRL05,OBrien_RIC1,OBrien_RIC3,OBrien_RIC2}.
This approach provides a lot of flexibility in controlling the amplitude, width, and frequency of the absorption resonances for the idler (Stokes) field by adjusting the parameters of the strong Raman control field $\Omega_A$.
The resonant susceptibility of this alternative $\Lambda$-system is:
\begin{equation}
 \chi_{\text{abs}} =  \frac{3\gamma_{r}^{\text{s2}}N_{\text{s2}}\lambda_{\text{s2}}^3}{8\pi^2} \frac{|\Omega_A|^2 \left(\Delta_{2} +i \gamma_{\text{ac}}\right)^{-1}}{ \left(\delta_{2} +\Delta_{2} -i \gamma_{\text{ab}}\right) \left(\delta_{2} -i \gamma_{\text{bc}}\right)-|\Omega_A|^2}.
\end{equation}
where $\Omega_A$ and $\Delta_2$ are the Rabi frequency and the applied Raman control field with detuning and its detuning from the optical transition $|a\rangle - |c\rangle$, as shown in Fig.\ref{fig:rb_levels}, the radiative decay rate of the excited state $|a\rangle$ is $\gamma^{\text{s2}}_r$, and this second system has wavelength $\lambda_{\text{s2}}$ and density $N_{\text{s2}}$.
The resonant two-photon susceptibility for such scheme was derived in \cite{OBrien_RIC1}:
\begin{align}
 \chi_{\text{2ph}} =  \frac{3N_{\text{s2}}\lambda_{\text{s2}}^3}{8\pi^2} \frac{|\Omega_A|^2}{\Delta_2^2} \frac{\gamma_{r}^{\text{s2}}}{|\Omega_A|^2/\Delta_2 +i(\gamma_{cb} + \gamma _{ab} \frac{|\Omega_A|^2}{\Delta_2 ^2})}.
\end{align}
It is easy to see that the effective optical depth due to the two-photon absorption is:
\begin{align}
D_{\text{abs}} = \frac{\gamma_{ab}}{\gamma_{cb} + \gamma_{ab} \frac{|\Omega_A|^2}{\Delta_2 ^2}} \frac{|\Omega_A|^2}{\Delta _2 ^2} D_{2\text{L}}, \label{eq:14}
\end{align}
where $D_{2\text{L}} = 3\gamma_{r}^{\text{s2}}N_{\text{s2}}\lambda_{\text{s2}}^3/(8\pi^2 \gamma_{ab})$ is the peak optical depth for the corresponding two level system.
If the $\gamma _{cb}$ is small enough, it is possible to achieve the level of absorption from our effective three-level system as high as that of a bare two level absorption.

\begin{figure}
    \centering
    \includegraphics[width=\textwidth]{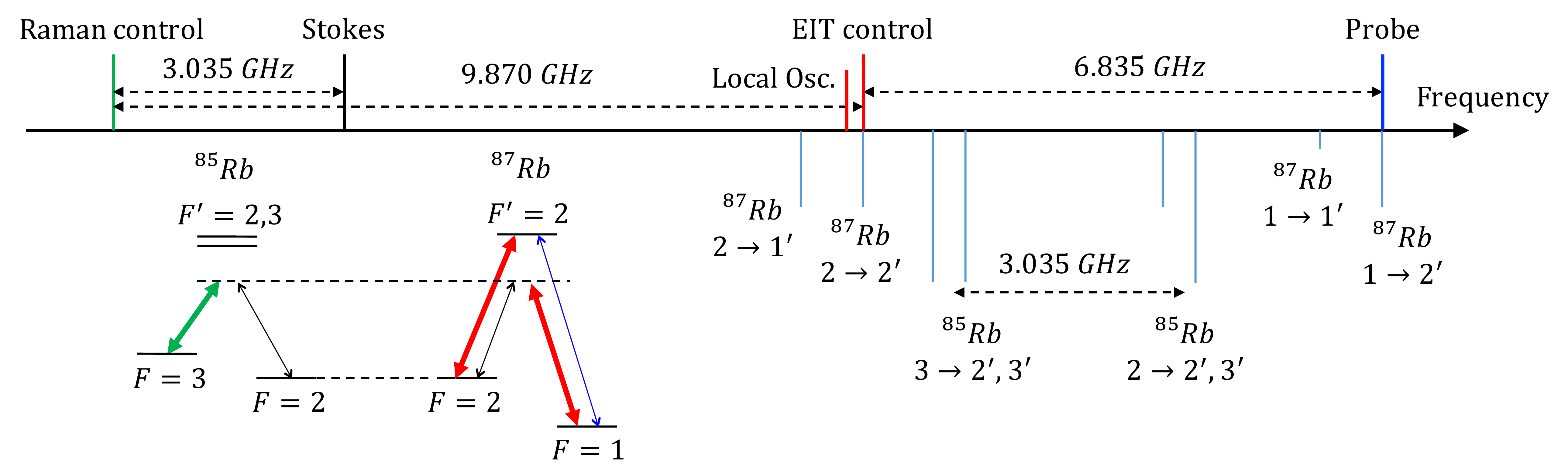}
    \caption{Spectral diagram of laser fields, involved in the experiments using two Rb isotopes. The color scheme is the same as in Fig.~\ref{fig:rb_levels}. Energy levels are not to scale.
    }
    \label{fig:laser_freq}
\end{figure}

The most natural implementation of this idea is in Rubidium gas, since it has two available isotopes with closely matched transitions.
In our experiments, we chose to implement EIT in ${}^{87}$Rb atoms, and Raman absorption in ${}^{85}$Rb atoms. This configuration was a natural choice, since the higher abundance of ${}^{85}$Rb (72\%) compared to ${}^{87}$Rb (28\%) was expected to help ensure high enough optical depth for the Raman absorption.
The interaction arrangement, used in our experiment, is shown in Fig.~\ref{fig:laser_freq}. For EIT/FWM interaction the strong control field was tuned to the $5S_{1/2} F=2 \rightarrow 5P_{1/2} F'=1$ optical transition of ${}^{87}$Rb atoms.
The signal field's frequency was shifted by $\Delta_{\mathrm{87}}\simeq 6.835$~GHz to the blue, to match the frequency of the $5S_{1/2} F=1 \rightarrow 5P_{1/2} F'=1$ transition.
We also sent in a non-zero input Stokes field, detuned by $\Delta_{\mathrm{87}}$ to the red from the control field, fulfilling the FWM resonance conditions.  

The Raman absorption resonance at the Stokes field frequency was created by another strong control field, shifted by  $\Delta_{\mathrm{85}}\simeq 3.035$~GHz to the blue from the Stokes field.
This way, it formed a far-detuned $\Lambda$ system, based on the $5S_{1/2} F=2,3 \rightarrow 5P_{1/2} F'$ transitions of the ${}^{85}$Rb atoms.
The exact frequency of this Raman control field was adjusted to maximize the absorption of the Stokes field, which was convenient to do with the seeded field.

\begin{figure}
   \centering
   \begin{subfigure}[b]{0.3\textwidth}
       \includegraphics[width=\textwidth]{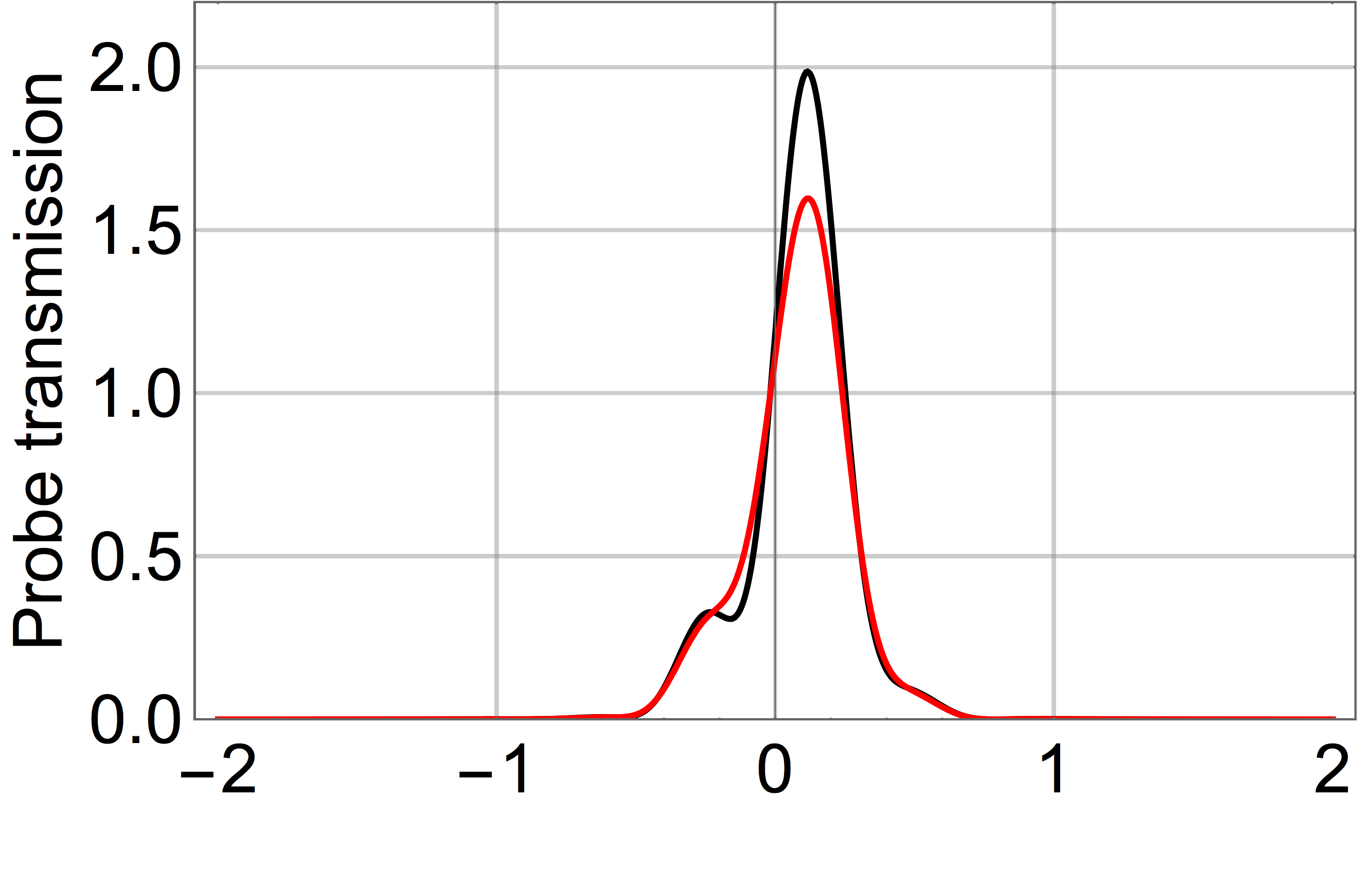}
   \end{subfigure}
   \begin{subfigure}[b]{0.3\textwidth}
       \includegraphics[width=\textwidth]{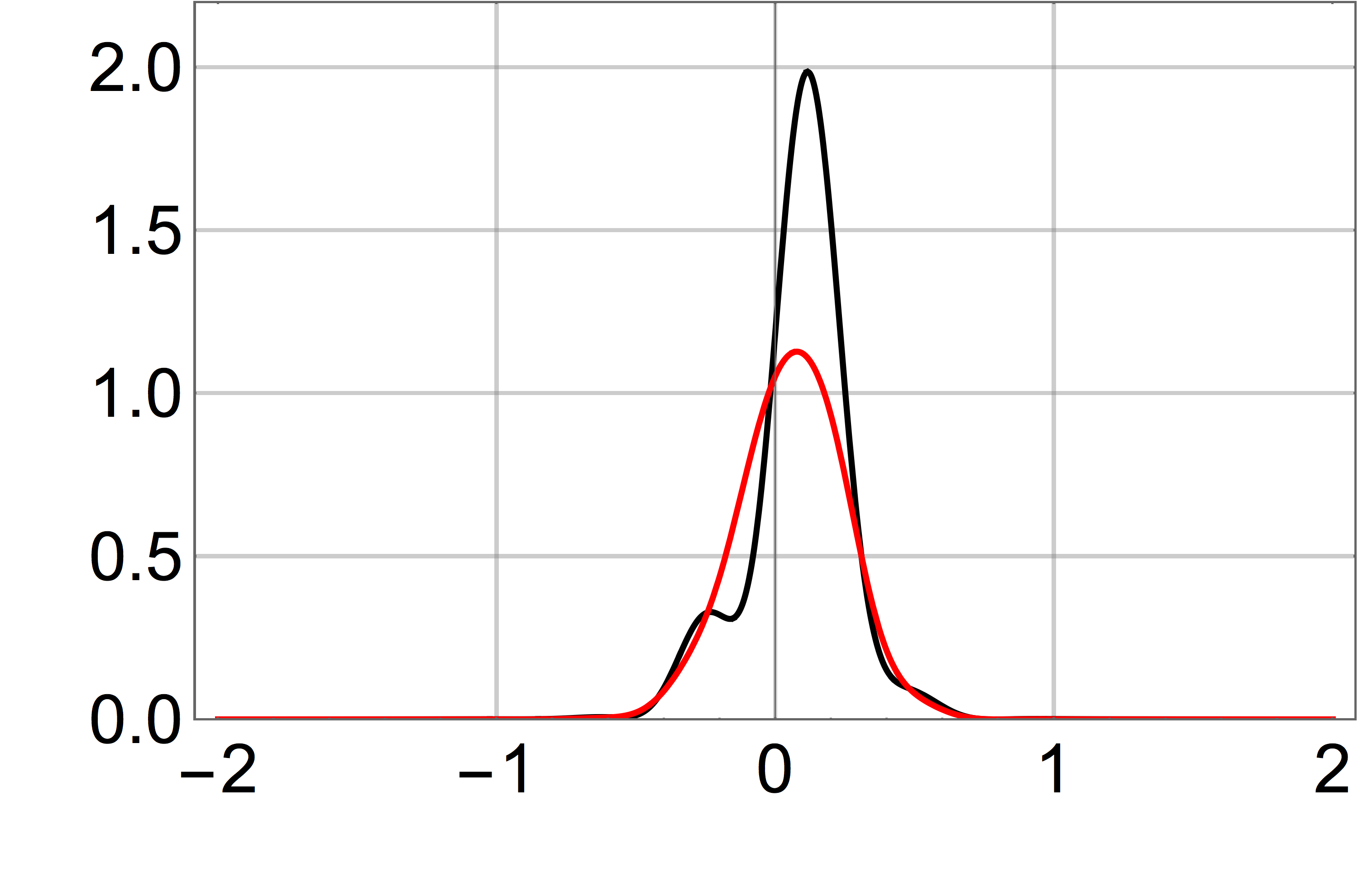}
         \end{subfigure}
 \begin{subfigure}[b]{0.3\textwidth}
       \includegraphics[width=\textwidth]{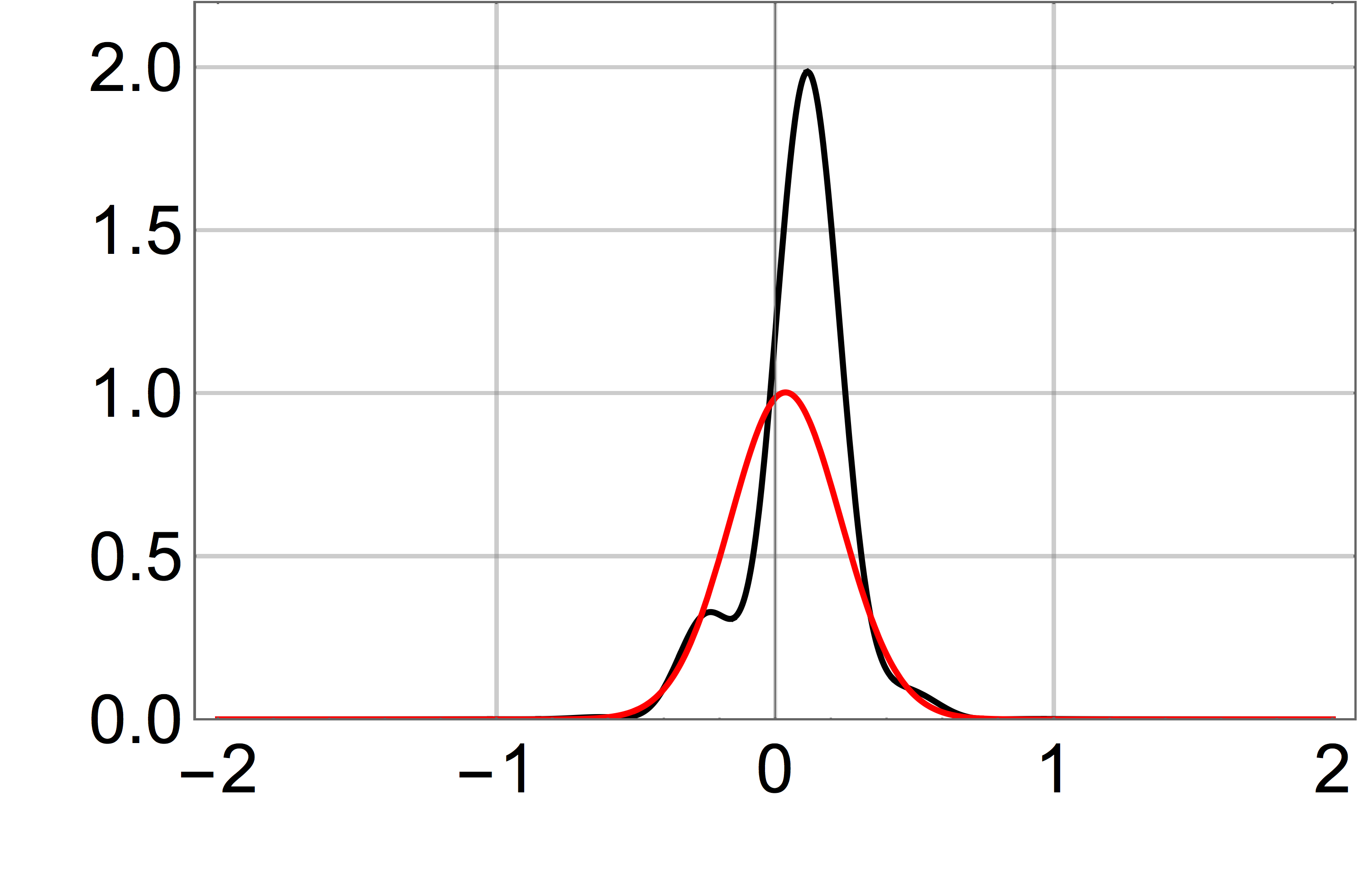}
          \end{subfigure}       
         
   \begin{subfigure}[b]{0.3\textwidth}
       \includegraphics[width=\textwidth]{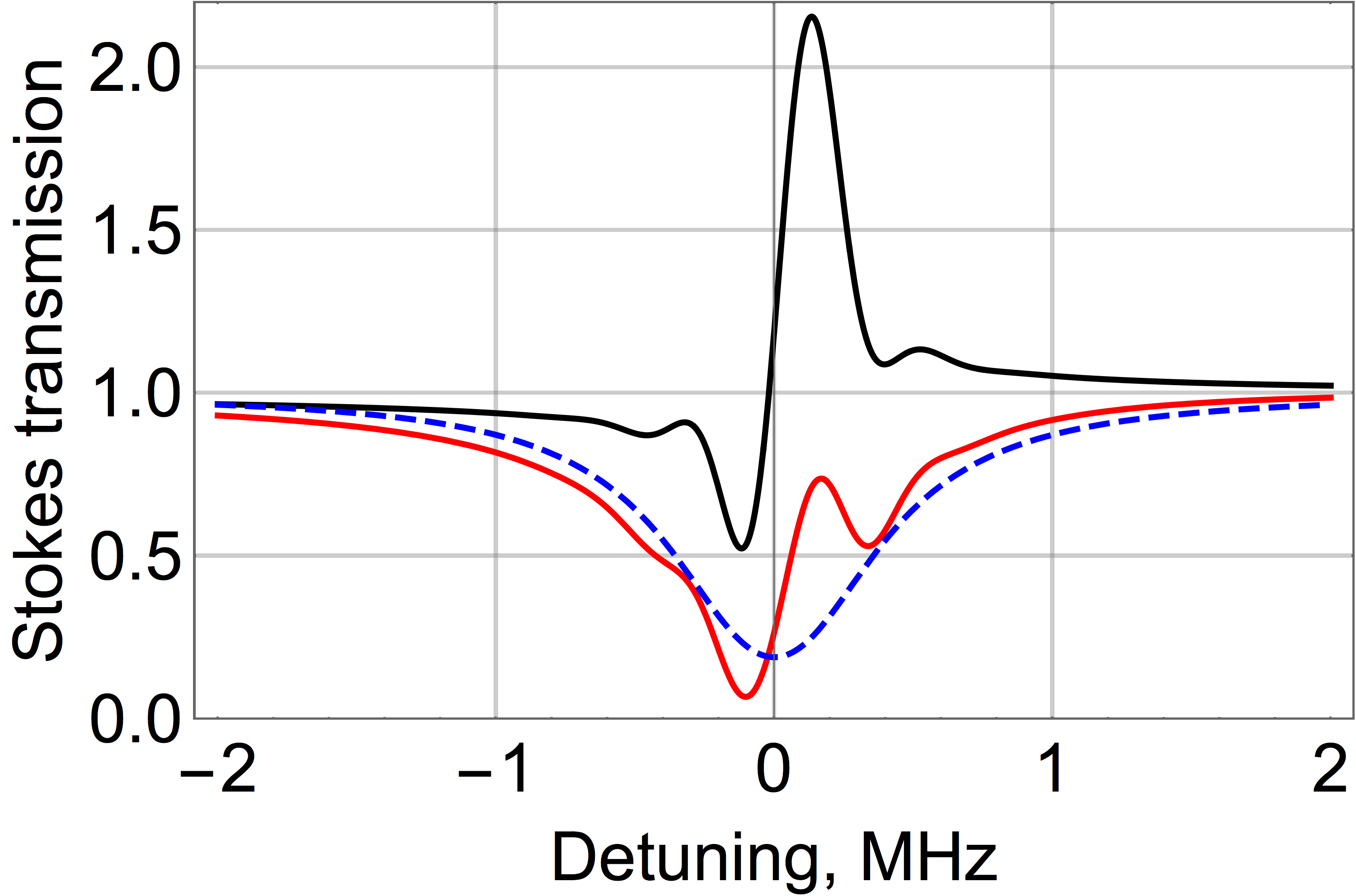}
          \end{subfigure}
      \begin{subfigure}[b]{0.3\textwidth}
       \includegraphics[width=\textwidth]{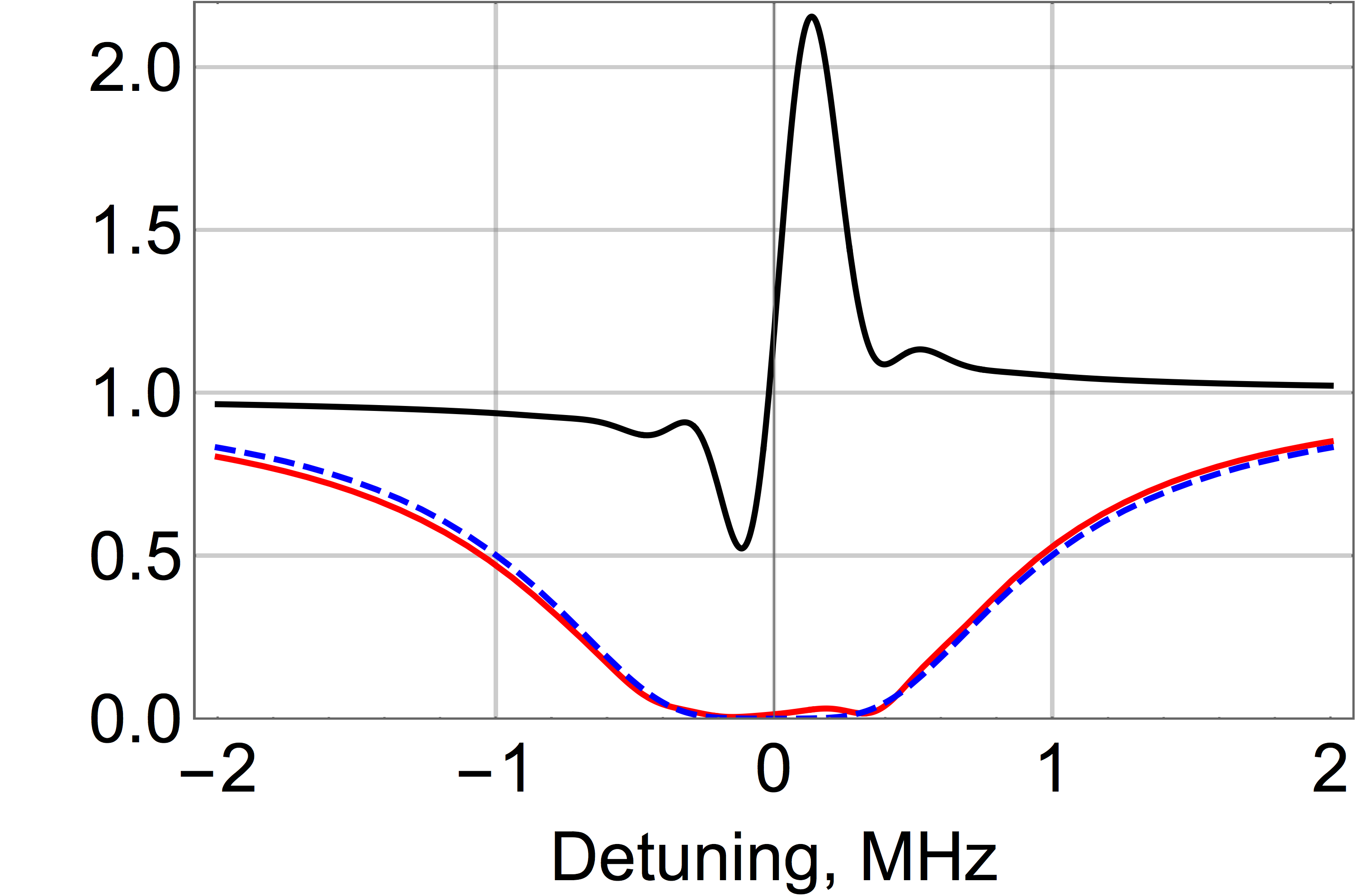}
         \end{subfigure}
   \begin{subfigure}[b]{0.3\textwidth}
       \includegraphics[width=\textwidth]{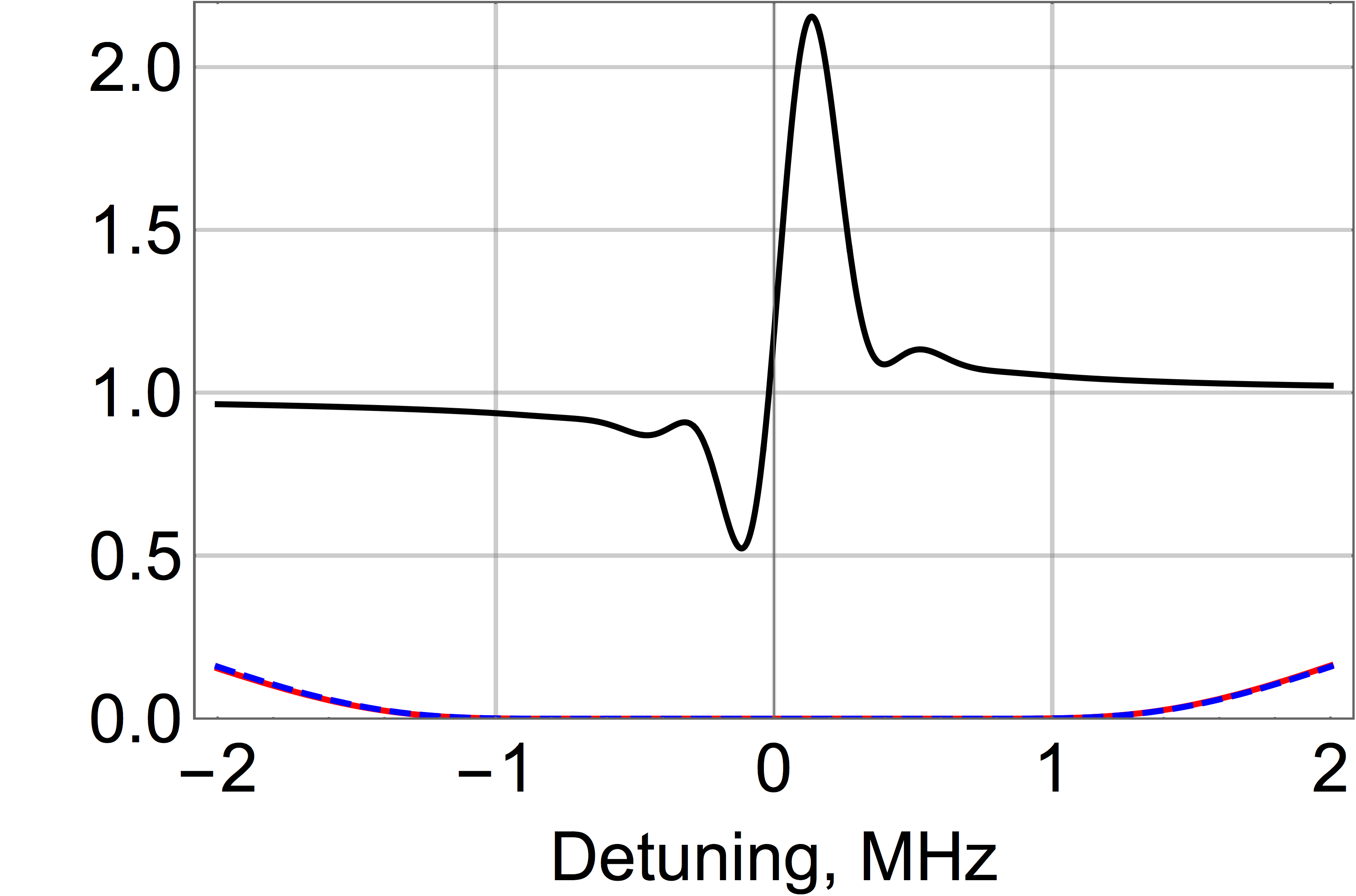}
          \end{subfigure}
   \caption{Probe (top row) and Stokes (bottom row) transmission as a function of the two-photon detuning under the FWM conditions for $ D_{abs}=0.83 $ (left column), $ D_{abs}=4.16 $  (middle column) and $ D_{abs}=41.6 $  (right column).
   Black curves are the resonances without the Raman absorption, red curves are the resonances with the Raman absorption turned on and the blue dashed curves are the shapes of the applied absorber.}
   \label{fig:fwm_reduction_1}
\end{figure}

Fig.~\ref{fig:fwm_reduction_1} illustrate the evolution of the EIT and FWM resonances as the effective optical depth of the Raman absorption increases. The reference black lines on each graph represent the lineshape of either probe or Stokes field transmission with no additional Stokes absorption. In this case we observe a strong amplification of either signals, but also the modification of the resonance lineshape from a traditional Loretzian lineshape. Such distortion, observed previously in the experiments, can be explained by the effective interference of the two-photon EIT channel and the four-photon FWM channel, each of which modifies the amplitude and the phase of the participating optical fields~\cite{hong09,phillipsJMO09,phillipsPRA11}. Thus, for different values of the two-photon detuning the contributions from these two channels add up either constructively or destructively. Such lineshape modification can be used to indirectly characterize the relative strength of the four-wave mixing process.

In the presence of the Raman resonance, the amplification of either Stokes or probe is reduced. However, if the absorption resonance is not strong enough [Fig.~\ref{fig:fwm_reduction_1}(a,b)], the strongest suppression happens only near the bottom of the Raman resonance, and on its wings there is no dramatic change in the four-wave mixing effect. This observation emphasizes that another important parameter of the Raman resonance to consider is its spectral width. Clearly, the absorption resonance  should be broad enough matched to the bandwidth of the Stokes field generation, to ensure the effective FWM suppression over its entire spectrum. We can see that with increased strength of the Raman absorption all the output Stokes field can be efficiently suppressed, and the probe transmission approaches the symmetric EIT resonance with near-perfect resonant transmission, as it is expected for an ideal EIT three-level system. 

It is necessary to point out, however, that the theoretical calculations above included only an isolated Raman resonance, without full consideration for either effects of the Raman field on the optical transitions other than Raman transition, or for the residual resonant absorption of the Rb atoms, involved in the formation of the Raman transition. In practice, these effects played crucial role in experimentally observed performance.  The level configuration, depicted in Fig.~\ref{fig:laser_freq} was the most successful realization of the theoretical proposal for the natural abundance Rb vapor cell. We also tested the configuration where the Raman control was acting on the $ F=2 \rightarrow F'=2,3 $ transition in $^{85}$Rb, and instead of reduction of the Stokes field we observed additional four-wave mixing gain at the frequency of the Stokes field due to the additional strong field tuned close to the atomic transitions (see discussion in Section~\ref{discussion}).
We also tried switching the isotopes around and use $^{85}$Rb for EIT; however, in this case the low atomic density of the $^{85}$Rb atoms was not sufficient to create sufficient Raman absorption(maximum observed absorption resonance had an amplitude of only $3\%$). 

\section{Experimental setup}\label{expsetup}

In the experiment we used two lasers: an external cavity diode laser (ECDL) and a cw Ti:Sapphire laser.
Single-mode fibers (not shown) in the output of each laser ensured high quality Gaussian transverse intensity profile.
The ECDL served as a source of all optical fields for EIT/FWM measurements (control,probe and Stokes fields, as well as a local oscillator for detection).
Part of the beam was sent through an electro-optic modulator (EOM), that phase-modulated it at the frequency of the hyperfine splitting in $^{87}$Rb.
The $+1$ order modulation sideband was used as a probe field, and the $-1$ order modulation sideband was used as a seeded Stokes field.
After the EOM the beam passed through an acousto-optic modulator (AOM) which shifted the frequencies of all three fields by $+80$~MHz. Before entering the Rb vapor cell, the laser beams were collimated to the  diameter of approximately $1$~mm, and circularly polarized.
In this experiments we used a glass cell containing natural abundance Rubidium and 5 Torr of Helium buffer gas, placed inside a 3-layer magnetic shield with a heater around the innermost layer.
The cell length was 5~cm and its diameter was 2.5~cm.
The magnetic shielding greatly reduced the stray magnetic fields and the heater allowed us to regulate the atomic density of Rubidium. 

The ECDL was phase-locked to the Ti:Sapphire laser,  which was used as a control for the Raman absorption resonance.
The Raman control was 9.870~GHz red-detuned from the EIT control, 3.035~GHz red-detuned from the Stokes field and it was linearly polarized.

For the measurements the temperature of the cell was set to either $ 100~^{\circ}C $ or $ 90~^{\circ}C $, that corresponds to the densities of ${}^{87}$Rb atoms of $N_{87} \approx 1.3\cdot10^{12}$cm$^{-3}$ or $N_{87} \approx 0.7\cdot10^{12}$cm$^{-3}$, and the densities of ${}^{85}$Rb atoms of $N_{85} \approx 3.4\cdot10^{12}$cm$^{-3}$ or $N_{85} \approx 1.7\cdot10^{12}$cm$^{-3}$.
Laser intensities used in the experimental data shown below are:
\begin{center}
\begin{tabular}{ | l | r | }
   \hline
   EIT probe & 50~$\mu$W \\ \hline
   EIT control & 17~mW \\ \hline
   Stokes seed & 50~$\mu$W \\ \hline
   Raman control & 65~mW \\
   \hline
 \end{tabular}
\end{center}

\begin{figure}
    \centering
    \includegraphics[width=0.95\textwidth]{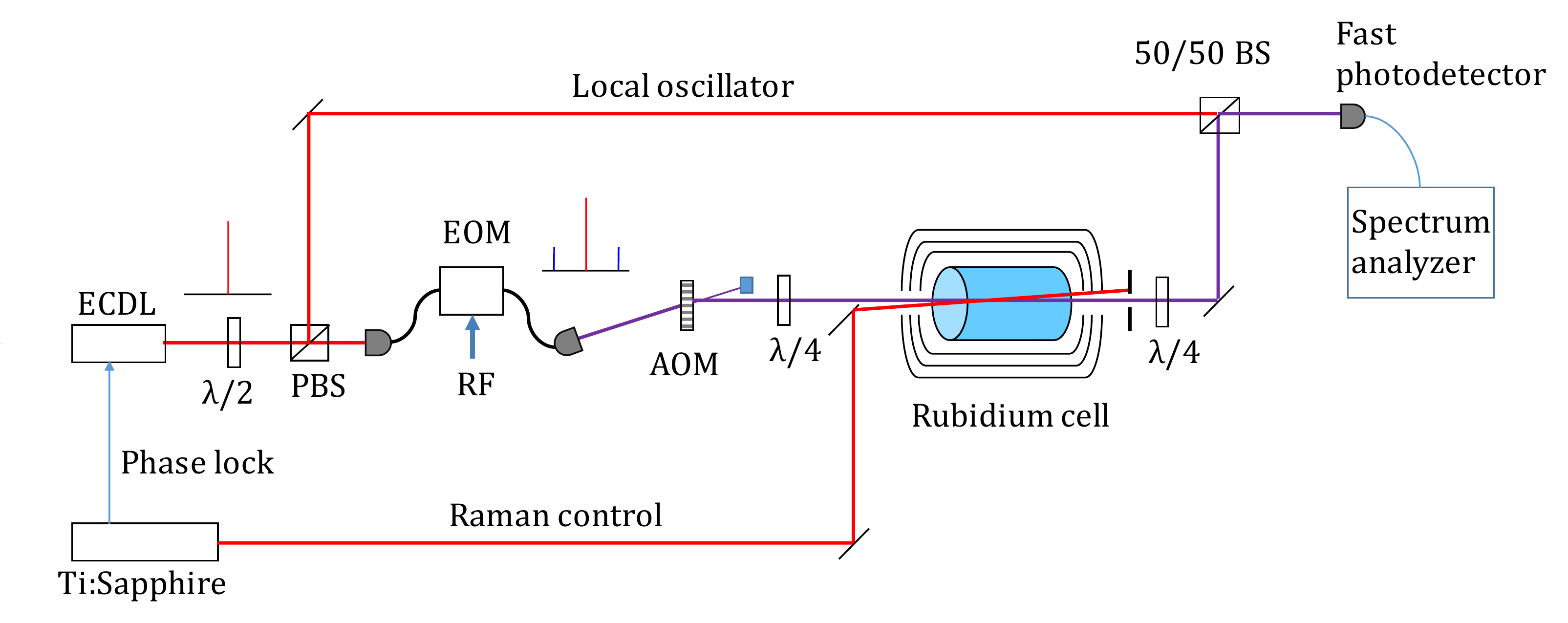}
    \caption{Schematics of the experimental setup. Here ECDL and Ti:Sapph are the External Cavity Diode laser and a cw Ti:Sapph laser, used in the experiment, $ \lambda/2 $ or $ \lambda/4 $ are half-wave  or quarter-wave plates, EOM is an Electro-Optic Modulator, AOM is an Acousto-Optic Modulator, and a  50/50 BS represents a 50/50 non-polarizing beam splitter.}
    \label{fig:exp_setup}
\end{figure}

We used a heterodyne detection scheme to record separately the transmission of the probe and Stokes optical fields. A part of the ECDL output was split off on a polarizing beam splitter before the AOM and used as a local oscillator.
It was combined with the main beam passing through the Rb vapor cell on a $50/50$ non-polarizing beam splitter, and the signal from the photo-detector was sent into a spectrum analyzer ( RBW~$=5$~MHz, VBW~$=3$~MHz, Span~$=0$~Hz ), where the beat-note signal between the local oscillator and the probe/Stokes field was detected. The probe signal was observed by setting the central frequency of the spectrum analyzer to $6.915$~GHz and the Stokes was at $6.755$~GHz.

\section{Experimental results}
\label{results}

\begin{figure*}[t!]
    \centering
    \begin{subfigure}[t]{0.5\textwidth}
        \centering
        \includegraphics[width=\textwidth]{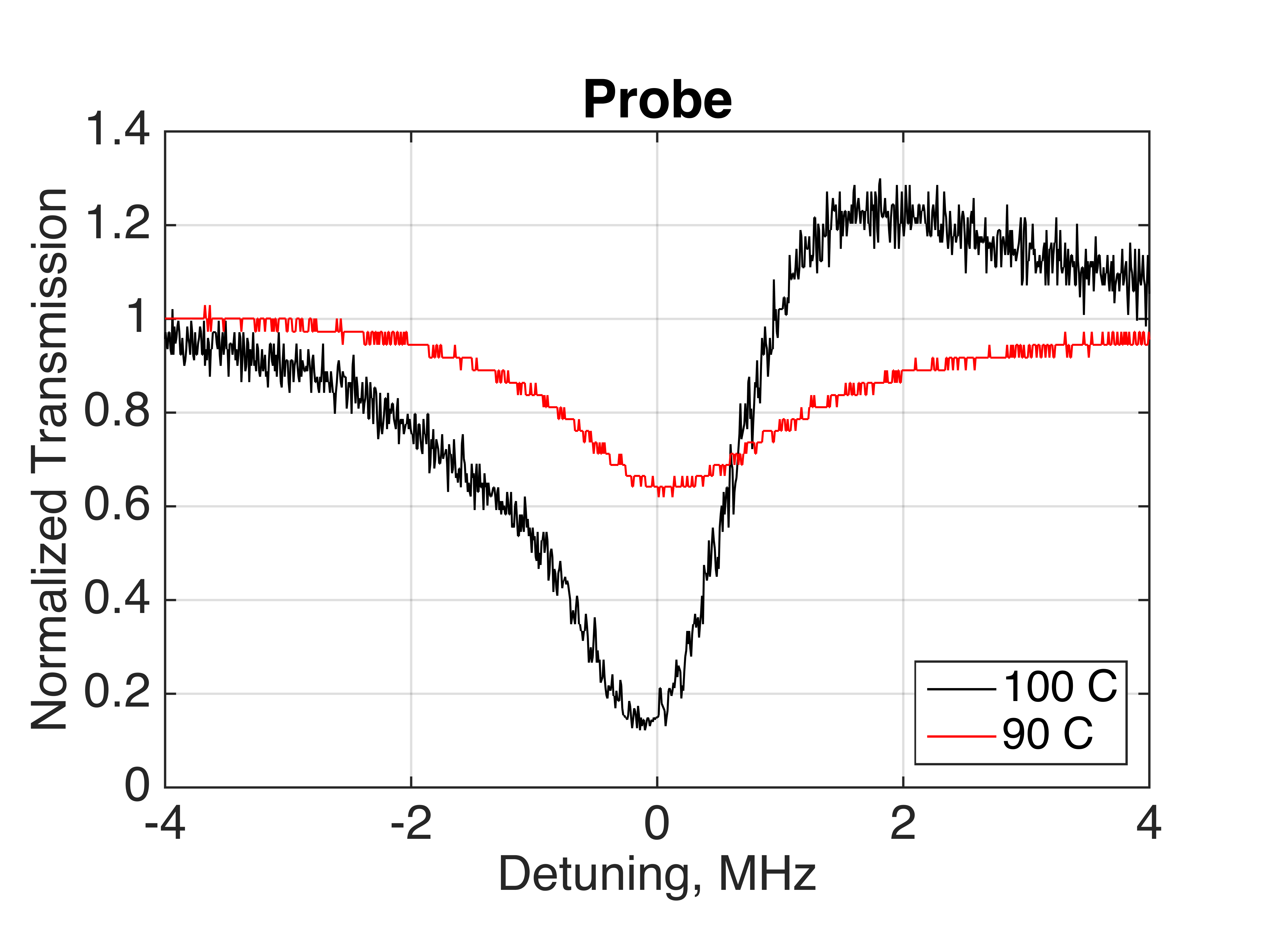}
        \caption{ }
    \end{subfigure}%
    ~ 
    \begin{subfigure}[t]{0.5\textwidth}
        \centering
        \includegraphics[width=\textwidth]{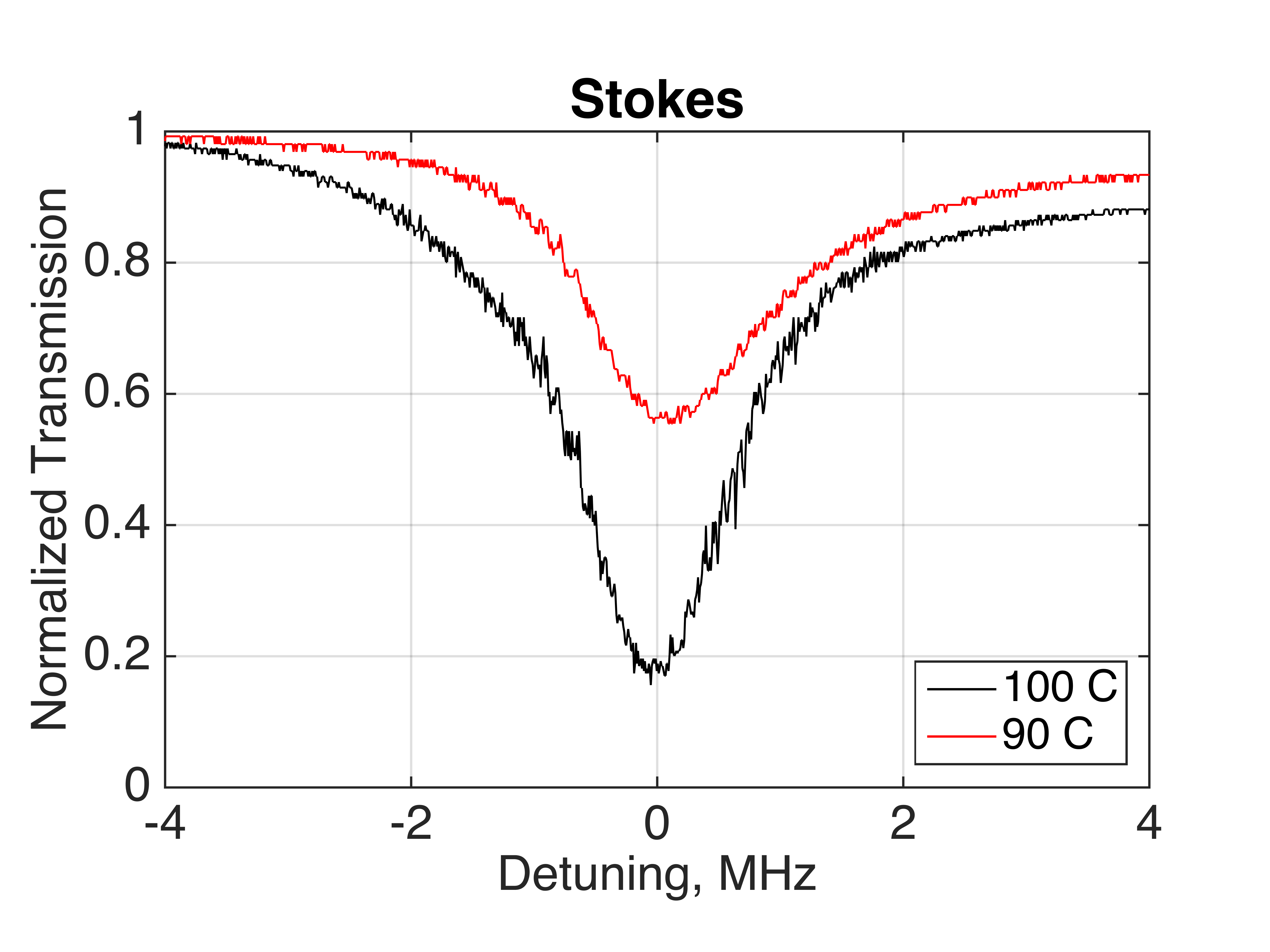}
        \caption{ }
    \end{subfigure}
    \caption{Raman absorption resonance for the probe field (a) and the Stokes field (b) as the Raman control .
    The traces were normalized to their corresponding transmission values, recorded with the Raman control field red-detuned by 5~MHz from the absorption resonance center.}
\label{fig:pl_scan}
\end{figure*}

We first characterized the absorption resonance, induced by the Raman control field, produced by the Ti:Sapphire laser, for two different cell temperatures -- 90 and 100 $^{\circ}C$. To do that we fix the EOM modulator frequency to maximize the probe transmission (6.835~GHz). The frequency of the Ti:Sapphire laser was tuned 3.035~GHz to the red of the Stokes field (and approximately 9.790~GHz to the red with respect to the ECDL laser frequency), as shown in Fig.~\ref{fig:laser_freq}, to find the absorption resonance, and then scanned in its vicinity to record the absorption profile.
The results are presented on Figure~\ref{fig:pl_scan}.

Looking at the relative change in the transmission for the two fields, we found that for the fixed Raman control power, the amplitude of the absorption peak increased with atomic density, reaching $ \approx 80\% $ at 100~$^{\circ}C$, as predicted by the simple theory. We also observed strong effect on the probe field transmission, in particular a significant drop at the Ti:Sapph frequencies, corresponding to the maximum Stokes field absorption. However, to fully appreciate this data, it is necessary to look more carefully in the character of EIT resonances.

\begin{figure*}[t!]
    \centering
    \begin{subfigure}[t]{0.5\textwidth}
        \centering
        \includegraphics[width=\textwidth]{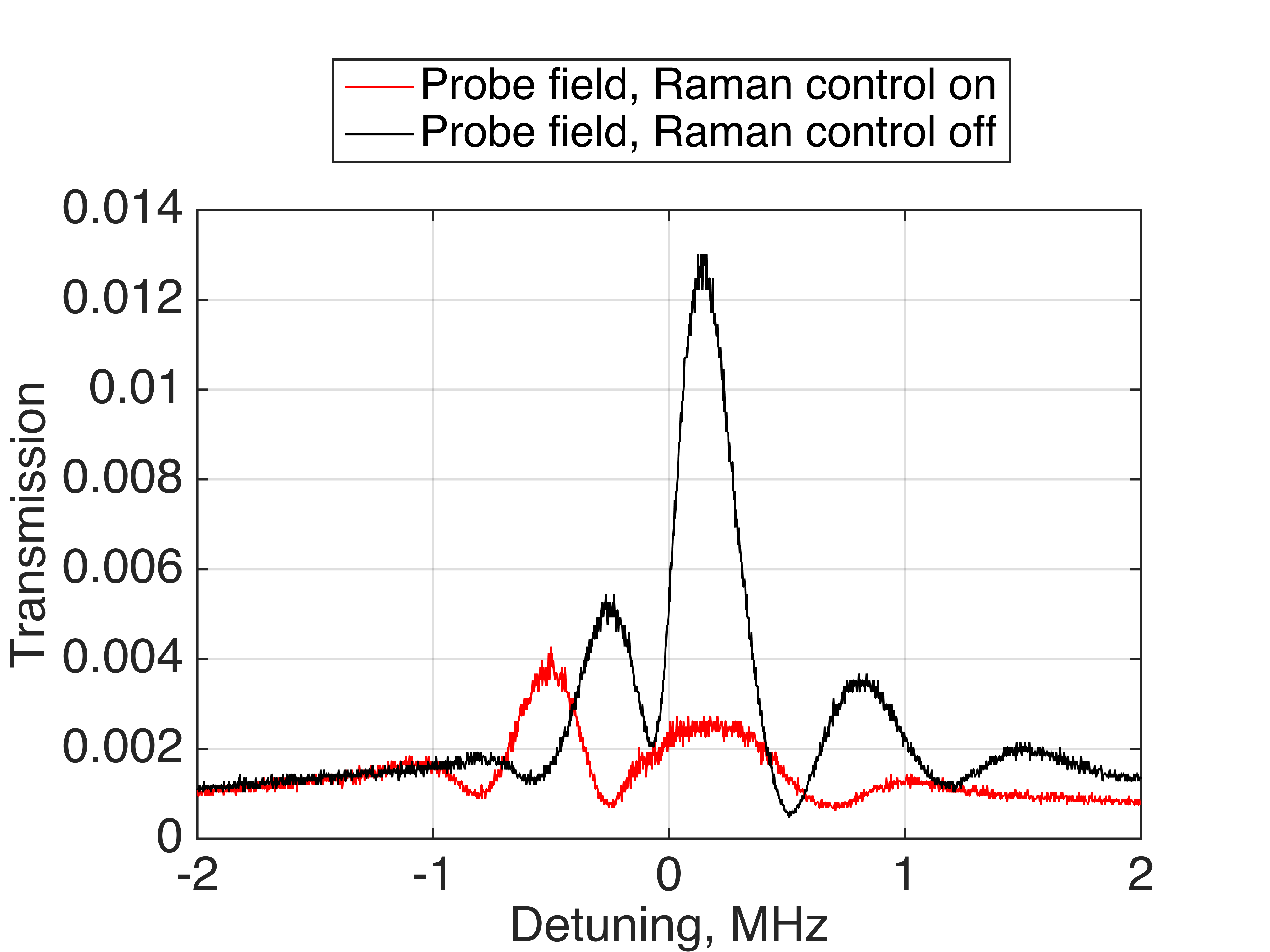}
        \caption{ }
    \end{subfigure}%
    ~ 
    \begin{subfigure}[t]{0.5\textwidth}
        \centering
        \includegraphics[width=\textwidth]{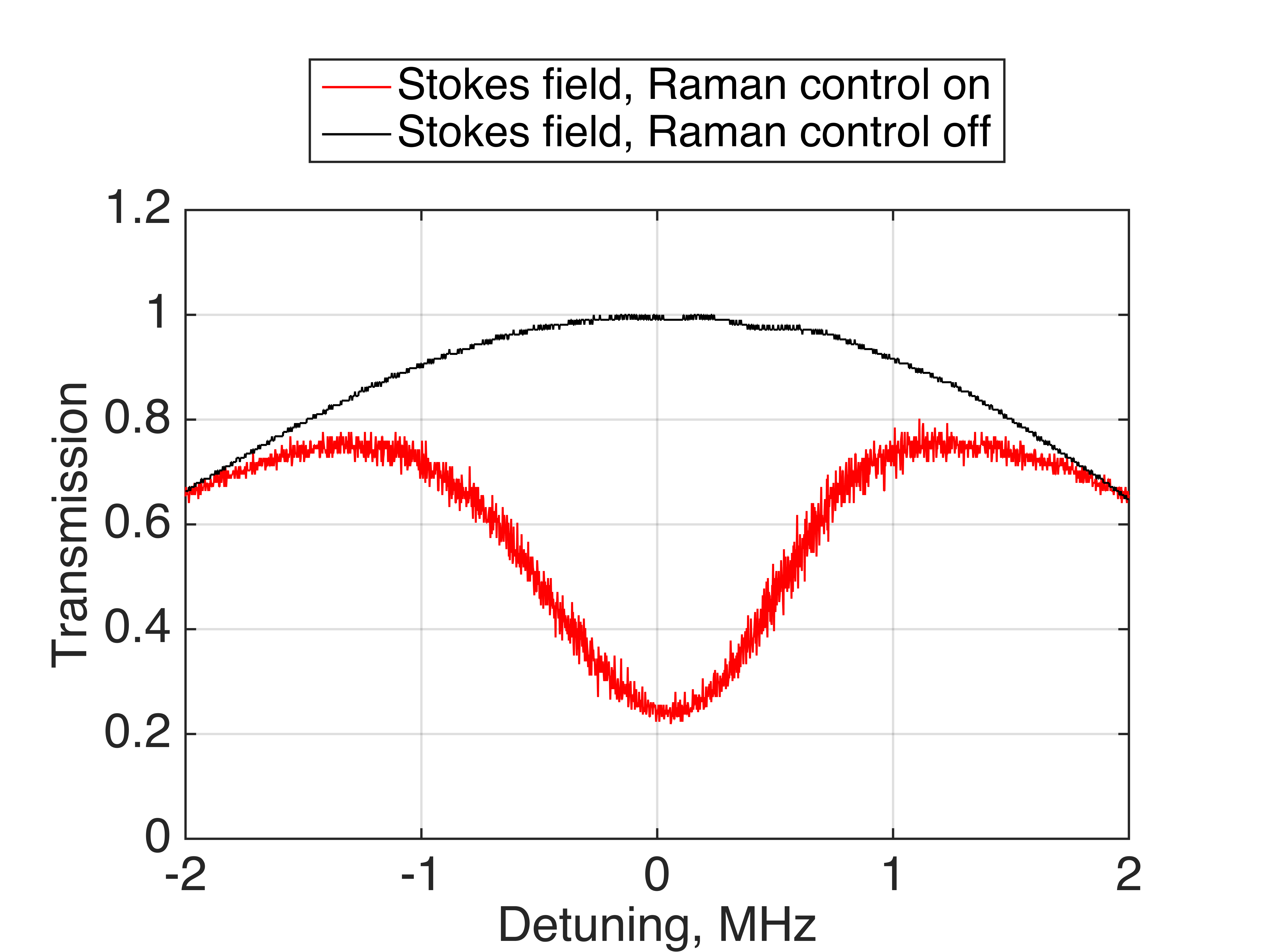}
        \caption{ }
    \end{subfigure}
    ~
        \begin{subfigure}[t]{0.5\textwidth}
        \centering
        \includegraphics[width=\textwidth]{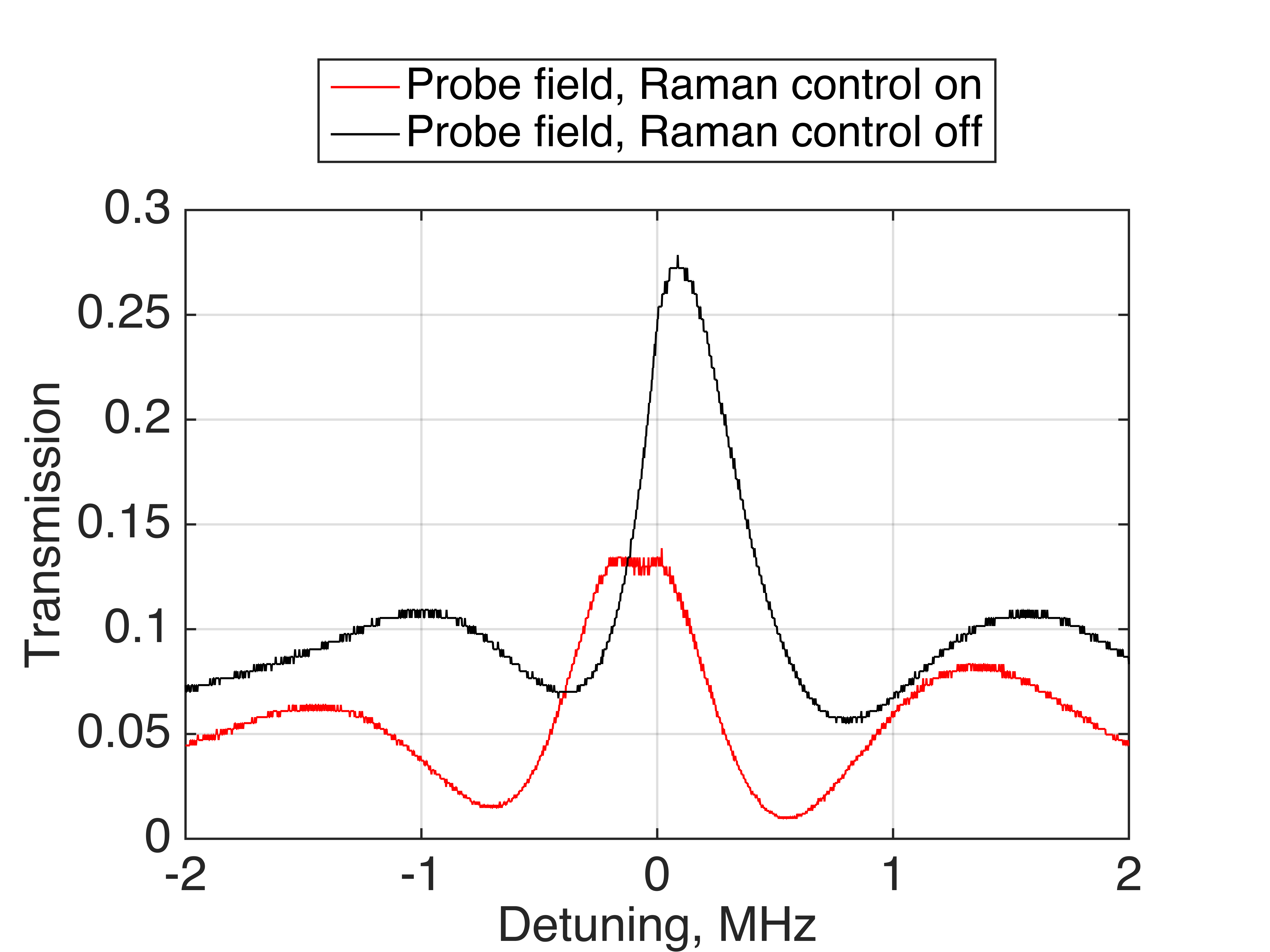}
        \caption{ }
    \end{subfigure}%
    ~ 
    \begin{subfigure}[t]{0.5\textwidth}
        \centering
        \includegraphics[width=\textwidth]{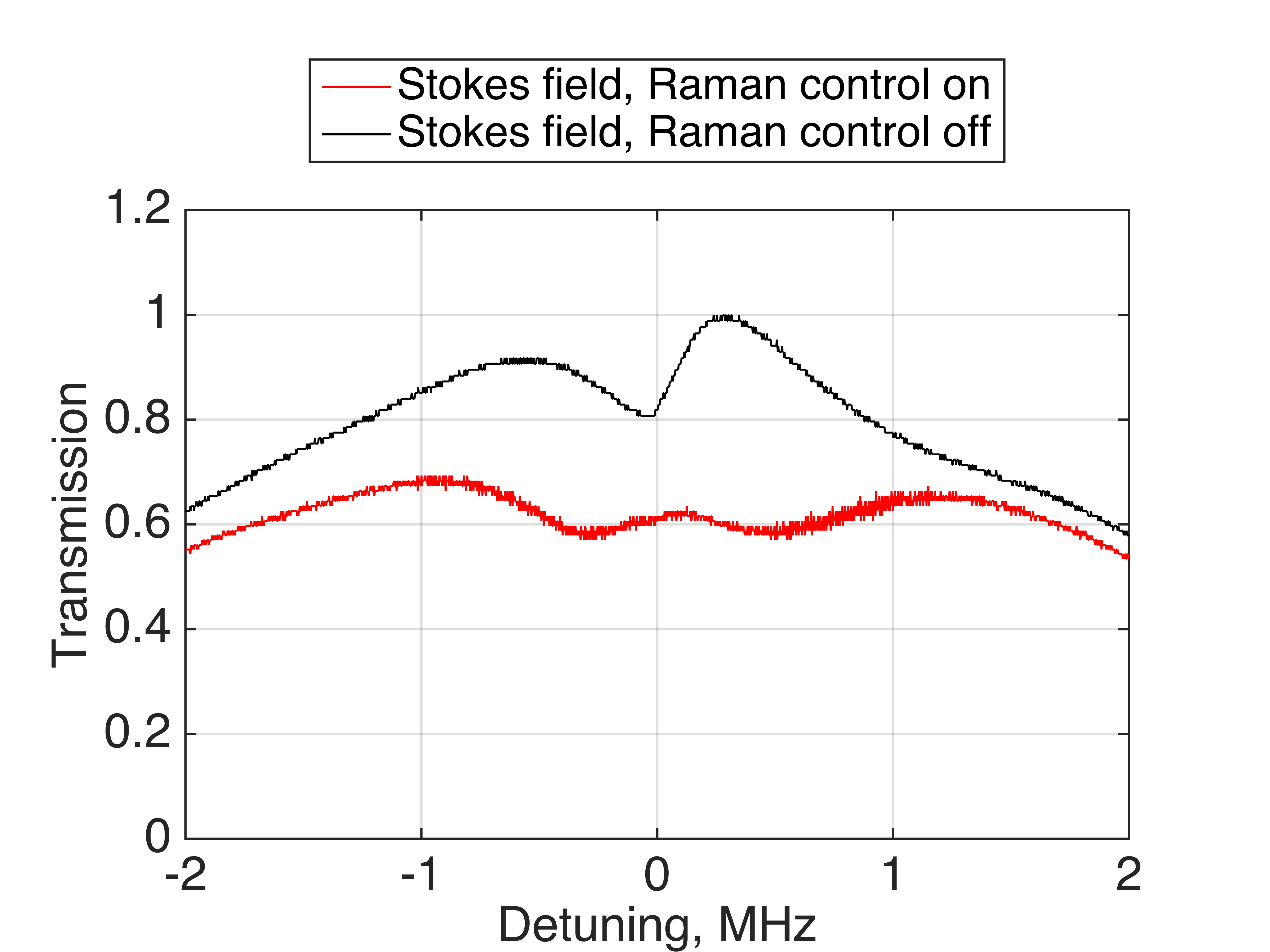}
        \caption{ }
    \end{subfigure}
    \caption{Transmission of the probe (a,c) and Stokes (b,d) fields through the cell versus two-photon detuning at 100~$^{\circ}C$ (a,b) and 90~$^{\circ}C$ (c,d).
    Here we scan the EOM modulation frequency around 6.835~GHz.
    The Raman control frequency is fixed to produce the absorption resonance at the center of the EIT resonance.
    During the scans the detection frequency of the spectrum analyzer remained fixed, which led to the roll-off on the plots due to the limited bandwidth of the spectrum analyzer.
    The traces were normalized to the maximum value of the Stokes field after the Rubidium cell.
    }
\label{fig:eit_scan}
\end{figure*}

To do that, we recorded probe and Stokes fields transmission resonances  for the same two cell temperatures as functions of their two-photon detuning.  
For these measurements the EOM modulation frequency was periodically scanned around the EIT transition frequency, simultaneously changing two-phton detuning of both probe and Stokes optical fields.
The Ti:Sapphire frequency was fixed at 9.790~GHz to the red with respect to the ECDL laser frequency and carefully adjusted so that the peak absorption coincided with the Stokes field.

The measurements of the probe and Stokes transmission without Raman control reveal a problem with our current experimental arrangement, caused by the strong absorption of the probe field by ${}^{85}$Rb atoms. Even though this field is detuned by a few GHz away from the ${}^{85}$Rb transition frequency, at high temperatures there is enough absorption at the wing of that optical resonance to seriously interfere with the probe propagation under EIT conditions. While for the lower temperature [Fig.~\ref{fig:eit_scan}(c)] the EIT contrast was approximately $30\%$, but for higher temperature [Fig.~\ref{fig:eit_scan}(a)] the overall height of the EIT transmission was only $<2\%$ of its input level. At the same time the multi-peaked shape of the EIT resonance indicate strong effect of four-photon four-wave mixing process, as discussed in Sec.~\ref{simpletheory}. It is logical to assume, however, that there is a significant four-wave mixing amplification present for the probe field, but its effect is countered by the strong absorption by the ${}^{85}$Rb atoms. Similar situation is with the seeded Stokes field: since its frequency is far away from any of the atomic resonances, it is transmitted largely without any modifications, although the effect of the four-wave mixing is visible at lower temperature [Fig.~\ref{fig:eit_scan} (b) ].

\begin{figure}
    \centering
    \includegraphics[width=0.5\textwidth]{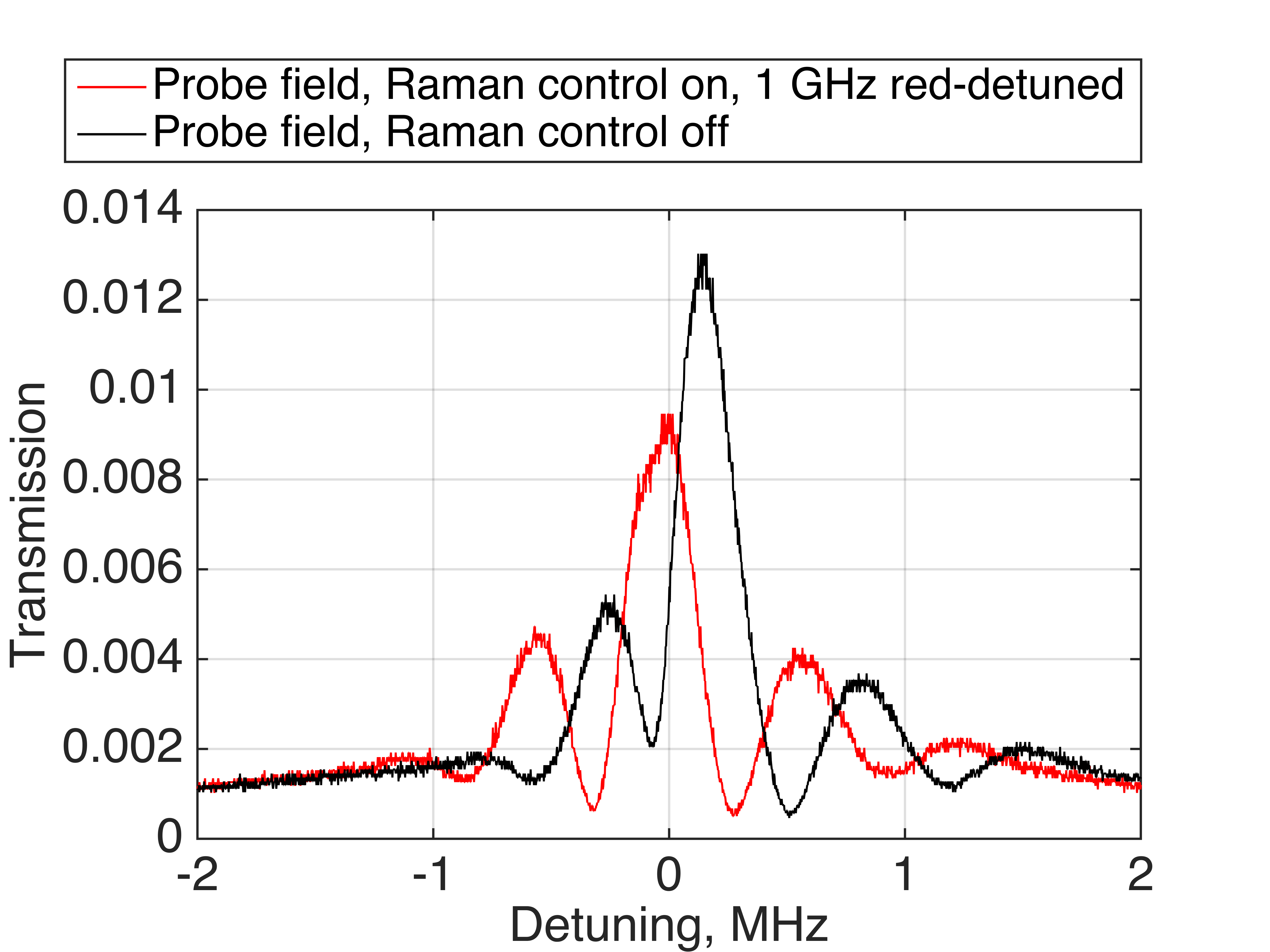}
    \caption{Transmission of the probe field through the cell versus two-photon detuning at 100~$^{\circ}C$.
    The Ti:Sapphire laser is 1~GHz red-detuned from the Raman absorption resonance.}
    \label{fig:eit_scan_probe_only_offset_100}
\end{figure}

With the Ti:Sapphire laser on, we observed an overall shift of the transmission resonance by $\approx 200$~kHz due to the light-shift from the additional control field.
We verified that by looking at the probe transmission when the Ti:Sapphire frequency was fixed at 10.790~GHz to the red with respect to the ECDL laser frequency, that corresponds to a 1~GHz red detuning with respect to the Raman absorption resonance, shown on Figure~\ref{fig:eit_scan_probe_only_offset_100}. In addition to such shift, however, we clearly observed the reduction of the probe transmission at frequencies of maximum Stokes absorption. This is expected for our case: even though our EIT control field is detuned to the EIT resonance, there is a strong probe absorption present in our atomic medium due to the second isotope. As a result, the output probe field is largely determined not by the transmission of the initial photons, but the four-wave mixing frequency conversion of the Stokes field~\cite{narducciPRA04}. However, if the four-wave mixing gain is suppressed, the probe output intensity rapidly falls.

\section{Further discussion}
\label{discussion}
From our preliminary experimental results it is clear that further optimization of the experimental parameters is necessary for successful realization of the proposed scheme. To achieve better transparency, we need the EIT control Rabi frequency needs to obey the following relationship
\begin{align}
\gamma _{ge} > |\Omega_C | \gg \sqrt{\gamma_{ge} \gamma_{gs}}.
\end{align}
In our experiment, the linewidth of the optical transition is dominated by Doppler broadening such that $\gamma_{ge} = 300$MHz.
The the ground-state spin decoherence rate is dominated by the rate at which atoms leave the illuminated area, and in our case is determined by the diffusion of Rb atoms in the presence of the 5~Torr of helium buffer gas. 
This rate can be estimated \cite{happer'72} giving a spin decoherence rate of $\gamma _{gs} = 64$~kHz, such that $\sqrt{\gamma_{ge} \gamma_{gs}} = 4.4$MHz. 
Thus, for our experimental conditions the optimal Rabi frequency should be around $\Omega _C = 50$MHz.
Because of the technical limitations of our EIT laser, we were not able to have high enough control laser intensities, as we can estimate that the experimental control laser Rabi frequency was $\approx 430$kHz, much too small for perfect EIT.   

However, higher control power is not the only parameter to consider. In principle, the Raman control field can also lead to spontaneouls Raman scattering in the original system, generating a second Stokes field at a different frequency, which also induces noise. 
Therefore, it is important to make sure the FWM strength of this new applied field is less than 1. 
Using Rb at natural abundance, this is actually impossible because the frequency of the applied control field for Raman gain is actually closer to resonance than the EIT control field.
To mitigate that effect, we need to switch the roles of the two isotopes, and to make sure that the density of ``Raman absorber'' isotope is higher than that of the ``EIT'' isotope in order to have the Raman absorption optical depth to be larger than the effective FWM optical depth. 
One solution would be to use an isotope mixture, for example, with 15\% ${}^{85}$Rb for EIT/storage and 85\% ${}^{87}$Rb for Raman absorption. In this case the EIT control field detuning is smaller $\Delta = 3.036$GHz, leading to stronger four-wave mixing signature~\cite{gengNJP2014}.

With this isotope mixture at the current experimental temperature, the optical depth for the probe field will still be high enough for efficient EIT storage ($D=15$), such that the FWM paramater in Eqs.(\ref{eq:1},\ref{eq:2}) is
\begin{align}
D \frac{\gamma _{ge}}{\Delta} = 1.48.
\end{align}
This indicates that the FWM effect is by far the dominant term, and a ``pure'' EIT quantum memory is impossible.
To create the required absorption resonance at the right frequency, the Raman control field, applied to ${}^{87}$Rb atoms, must be detuned by $\Delta _2 = 14.7$GHz  to match the Raman absorption to the generated Stokes frequency, as discussed in Sect.\ref{simpletheory}. 
Because of the chosen isotope ratio, the 2-level optical depth for ${}^{87}$Rb is $D_{2L} = (85/15)D = 85$, making it easier to achieve large enough Raman absorption depth to overcome the four-wave mixing gain.  
Plugging the above values into Eq.(\ref{eq:14}), we see that to have $D_{\text{abs}} = 1.1D$, one needs to have $|\Omega_A|^2/\Delta_2^2 = 5\cdot10^{-5}$, that can be achieved  with $\Omega_A \simeq 100$~MHz. 

It is also important to compare the spectral width of the Raman absorption line with the spectral width of the generated Stokes field. 
To fully suppress the four-wave mixing effects, the Raman absorption bandwith should exceed that of the EIT/FWM process, which requires~\cite{laukPRA13, OBrien_RIC1}:
\begin{align}
\gamma _{ab} \frac{|\Omega_A|^2}{\Delta_2^2} + \gamma _{cb} \left(1-\frac{|\Omega_A|^2}{\Delta_2^2}  \right) > \frac{|\Omega_C|^2}{\gamma_{ge} \sqrt{D}} \sqrt{\frac{2}{1+D/12}}.
\end{align}
With $\Omega_A = 100$MHz and $\Omega_C = 50$MHz, the Raman absorption width would be approximately $80$~kHz, which is a lot smaller than the Stokes field width of 680~kHz, estimated for the same experimental parameters.
If we can use a more intense control field for the Raman absorption the two can be matched if we take $\Omega_A = 700$~MHz. However, this more intense field will lead to larger rate of spontaneous Raman scattering, when interacting with EIT system. So under realistic conditions, we will likely have to live with a smaller absorption width, which means only absorbing the center of the Stokes line. At the same time, with strong absorption, the FWM will never begin and it should not matter that our absorption does not completely cover the full gain spectral bandwidth.  

On the other hand, if the Raman control field strength is kept at $\Omega_A = 100$MHz, then Eq.(\ref{eq:11}) gives: 
\begin{align}
\frac{N_{\text{abs}}}{N_{\text{FWM}}} \approx 5 \cdot 10^{-4},
\end{align}
predicting a complete elimination of the FWM noise. 

We also need to consider the spontaneous Raman scattering, induced by $\Omega_A$, which relative strength $x$ can be estimated as:
\begin{align}
x = D\frac{|\Omega_A|}{|\Omega_C|} \frac{\gamma_{ge}}{\Delta_A}.
\end{align}
With our experimental parameters, where $\Delta_A = 14.677$GHz is the detuning from resonance of the extra control field when applied to the ${}^{87}$Rb ground state, this process has an effective strength of $x = 0.64$, which is negligible. 

Comparing these numbers with the parameters of the experiment, it is easy to see that the main reason that FWM is not completely eliminated in the experiment is the same reason that EIT was not perfect: the intensity of the control laser used create Raman gain was too weak.
This explains why the FWM was only partially suppressed, as well as why extra Raman scattering was not observed.

\section{Conclusion}

In this manuscript we discussed the possibility to reduce the effect of the four-wave mixing on a probe signal in a dense atomic media under electromagnetically induced transparency conditions by selectively absorbing the idler (Stokes) optical field, also participating in the four-wave mixing process. Our theoretical calculations demonstrate that strong resonant absorption of the Stokes field effectively suppresses the effect of the four-photon interaction process (FWM) without affecting the two-photon EIT coupling of the probe and control fields. To create such a tunable absorption resonance for the Stokes field, we proposed to create a Raman absorption peak using a different atomic isotope and an additional strong laser field. We tested this proposal using a natural abundance Rb atoms, using ${}^{87}$Rb for EIT and four-wave mixing, and ${}^{85}$Rb for two-photon absorption resonance for the Stokes field. While our experiment did not achieve good electromagnetically induced transparency due to residual probe absorption by the wings of  ${}^{85}$Rb optical resonance, we were able to demonstrate that the four-wave mixing can be partially suppressed. We also discussed the more optimal set of experimental parameters.
We thank Dr. David F. Phillips (CfA) for lending  the Rubidium cell.
This research was supported by AFOSR grant FA9550-13-1-0098.
Dr. O'Brien would like to acknowledge support from NSERC (Canada). 


\end{document}